\documentclass[aps,pra,showpacs,floatfix,twocolumn,amsmath,amssymb,superscriptaddress,showkeys]{revtex4-1}
\usepackage{graphicx}
\usepackage{color}
\usepackage{ulem}
\usepackage{bm}
\usepackage{hyperref}

\def\etal{\textit{et al. }}

\def\no{\noindent}

\begin{document}
\title{Radiation spectrum of systems with condensed light}
\author{Alex Kruchkov\\
Laboratory for Quantum Magnetism (LQM), Institute of Condensed Matter Physics, 
\\
\'Ecole Polytechnique F\'ed\'erale de Lausanne (EPFL), 
Station 3, CH-1015 Lausanne,
             Switzerland
\\
alex.kruchkov@epfl.ch}      

\date{\today}

\begin{abstract}
Experimental observation of Bose-Einstein condensation (BEC) of photons inside a microcavity  induced an extensive study of the phenomenon.
 Beyond the purely theoretical interest,  this phenomenon is believed to be used to create a novel source of light. 
The shape of radiation spectrum  is therefore the main characteristic  of the system with light BEC as an optical device. 
In the present paper we propose the phenomenological model to calculate the shape of  the radiation spectrum of the system.
\end{abstract}





\maketitle

\section{Introduction}

Bose-Einstein condensation (BEC) of light is a spectacular phenomenon that was  observed in a dye-filled optical cavity \cite{Klaers1,Klaers2}. By its nature, BEC is an entirely quantum event, when the lowest energy state is populated with a macroscopically large amount of particles with integer spin (bosons), - such as, for example, photons are \cite{Einstein}.  Historically, the BEC was predicted to occur for many-particle boson systems  by Einstein  \cite{Einstein}   after he had got familiar with the work on photon statistics of Bose  \cite{Bose}.
Surprisingly, the BEC of light is one of a few quantum phenomena that is observed for normal conditions (room temperatures, usual pressure) \cite{Klaers2}, instead of extravagant and sophisticated ultracold experiments when BECs of alkali atoms were observed \cite{1,2,3}. Beyond the pure theoretical interest in the new state of light (see e.g. \cite{Kruchkov,Kruchkov1,Sobyanin,Kirton} and references therein), BEC of light being known better could improve energy conversion efficiency in solar cells (see  \cite{Kruchkov1} ).

The main indicator of the light condensation in experiments is the occupancy of the energy level, corresponding to the lowest possible energy of thermalized photons in the system under study \cite{Klaers2}. After thermal equilibrium of photon gas was gained, and then a sufficient amount of photons were pumped into the system, Klaers \etal \cite{Klaers2} observed the abrupt peak in the radiation spectrum of the dye-filled microcavity. The position of this peak corresponds exactly to the so-called cutoff frequency in the dispersion of photons, which plays the role of an effective mass of a photon in it's dispersion law.

The measured shape of radiation spectrum witnesses the brand new phenomenon, -  the condensation of light,  - which was considered impossible in past.
On the other hand, the shape of radiation spectrum  is the main characteristic of the system as an optical device. If implemented in a  novel source of light,  one should possess the sufficient knowledge about all the main optical characteristics of the light BEC in order to manipulate and control possible up-coming devices.
However, until recently no-one explained the shape of the spectrum in details.

A suitable non-interacting model to describe BEC of photons in thermodynamical equilibrium with  in-cavity medium was developed by Kruchkov and Slyusarenko \cite{Kruchkov,Kruchkov1} and independently by Sobyanin \cite{Sobyanin}. Kirton and Keeling \cite{Kirton} developed a non-equilibrium, which, in particular gave the deeper insight on the differences between BEC of photons and conventional single-mode lasing. Recent experimental results of Schmitt et al. \cite{fluctuations} shows the underestimated importance of statistical fluctuations in the system under study. 
Interacting models for the system under study are currently restricted to the Gross-Pitaevskii equation or its modifications \cite{Klaers2,Nyman}. 
The further experimental success of Klaers and colleagues (see e.g. \cite{fluctuations,optomech}) induced a number of recent theoretical works \cite{statistical,Leeuw1,Leeuw2,2014.arxiv.1,2014.arxiv.2,2014.arxiv.3,2014.arxiv.4}.
 Yet there was the challenge to explain the shape of the BEC light spectrum. 
In the present paper we propose the phenomenological model which is in good correlation with experiments \cite{Klaers2}.

The structure of the paper is following: 
In Sec.II we draw attention to the specificities of the considered phenomenon, in particular discussing the differences from the experiments on Bose-Einstein condensation of ultracold atomic gases (see e.g. \cite{1,2,3}). In Sec.III we propose the phenomenological model to describe the shape of radiation spectrum, the origin of the thermal tail in the energy-forbidden part of the spectrum, also as the description of the thermal blur of the condensate peak, which was observed in the experiments. Finally, in Sec.IV we discuss the influence of photon-photon interactions in the framework of the nonlinear model used by the authors of Ref.\cite{Klaers2}.

\section{Specifics of light condensation experiments}

Let us first directly address  to the  the crucial differences from the traditional atomic BECs, by reason of which for a long time 
the idea of  experimental implementation of light condensates was considered to be preposterous, pending the moment of emergence of the iconoclastic papers of Klaers, Schmitt, Vewinger and Weitz in 2010 \cite{Klaers1,Klaers2}. In the present section we list five the most important, on our opinion, specificities of light condensation experiments, and ask  the reader to use  Refs.\cite{Klaers1,Klaers2,Kruchkov1, Sobyanin,Kirton} for additional details.

\no
(i)
According to Planck's law, photons vanish while cooling a thermodynamical system to the absolute zero of temperature. Thence, in contrast to experiments with atomic BECs, the cooling of the system do not ensure the Bose-Einstein phase transition of photons.

\no
(ii) 
In compliance with results of up-to-date experiments, photons are considered to be massless particles (quanta of electromagnetic field). 
Nevertheless, under definite conditions (e.g. placed in waveguides or cavities)  the photons can reveal the presence of an {\it{effective mass}} in their behavior.
In particular, the appearance of the effective mass in the considered system \cite{Klaers1,Klaers2} happens essentially due to the thermalization of the single longitudinal mode (see further). Note that  the presence of mass term in the dispersion relation is not, generally speaking, the necessary condition for the Bose-Einstein condensation, however this quantity enriches significantly the physical and mathematical analogy between light condensates and atomic BECs of the reduced dimensionality. 

\no
(iii)
Bose-Einstein condensation of light is a  non-equilibrium (in statistical sense) phenomenon. By reason of the considerable amount of absorption and scattering processes, also as a small uncontrollable escape of photons through the edges of the experimental setup \cite{Klaers2}, the total number of free  photons in the system is not conserved.  Therefore to maintain the finite chemical potential of photons, the system requires a weak periodical photon pumping followed by  relaxation processes. 
Thereby, the {\it{average}} number of photons is conserved, and we can talk about the condensation \cite{Klaers2,Kruchkov1,Sobyanin,Kirton}.

\no
(iv)
The complexity of thermalization processes \cite{Klaers2} does not allow to change the temperature of the system arbitrarily. In this connection, the temperature becomes only a property of the system, meanwhile the total number of free photons is the description parameter. 
As opposed to experiments on atomic BECs \cite{1,2,3}, it is not  the cooling  of the system to guarantee the Bose-Einstein condensation of photons, but the controllable increase in total amount of quantum particles. Such experimental strategy allows to observe BEC of photons even for comfortable room temperatures ($300$ K), that is nine order of magnitude higher than critical temperatures of alkali metals BECs (e.g. $170$  nK for Rubidium-87).

\no
(v) The success of the experiments on thermalization and condensation of photons in an optical microcavity depends significantly on the reasonable selection of parameters of the cavity and the filling medium (the organic dye in \cite{Klaers1,Klaers2}), and specifically on the correct and nontrivial choice of absorption and emission spectra taking into account energy levels of the cavity.

We now briefly remind the properties of photons inside a narrow cavity (see Fig. 1) with highly reflective walls\cite{Klaers1,Klaers2,Chiao,Sobyanin,Kruchkov1}. A photon as a relativistic object possess the dispersion relation $\hbar \omega = \hbar \tilde c k=  \hbar \tilde c \sqrt{k_z^2+\mathbf{k}_{\mathbf{r}}^2}$ (here $\tilde c$ is the speed of light in the medium). In the single-mode system, i.e. when the distance between $k_z$ is large and the only $k_z$ mode is thermalized \cite{Klaers2},
 light dispersion relation inside a cavity  can be presented in a form as if photons were two-dimensional (2D) non-relativistic particles with a 2D momentum $\mathbf{k}_{\mathbf{r}}$ and an effective mass $m^{*}$, corresponding to the cutoff energy $\hbar \omega_{0}= m^{*} \tilde c^{2} $  (see also \cite{Chiao}) . Constructing a cavity with spherically curved highly reflective mirrors, we actually put photons in a pure geometrical trapping potential  $\Omega$ (for details see  \cite{Klaers2,Kruchkov1,Sobyanin}):

\begin{equation}\begin{split}\begin{gathered}
\label{dispersion}
\hbar  \tilde{\omega} \left( k_{r},r \right) 
\approx
 \hbar {\omega _0} + \frac{{{\hbar ^2}k_r^2}}{{2{m^*}}} + \frac{1}{2}{m^*}{\Omega ^2}{r^2}.
\end{gathered}\end{split}\end{equation}

\no
Note that the cutoff frequency $\omega_0$ (and, correspondingly, the effective mass of a photon) rises from the boundary conditions to the Maxwell equations for electromagnetic field inside a cavity; in particular, in case of a narrow  cavity with highly reflective walls, the cutoff frequency with a good accuracy is  $\omega_0 = \pi \tilde c q / {l_0}$, where $q$ is a longitudinal mode number (an integer) and $l_0$ is the width of the cavity at the symmetry axis (for details see e.g. \cite{Klaers1,Klaers2,Kruchkov1,Sobyanin}). As we have already mentioned, the uniqueness of $\omega_0$ is provided by the thermalization of a single longitudinal mode, for example $q=7$ (for details see \cite{Klaers1,Klaers2}).

\begin{figure}[t]
\center{\includegraphics[width=0.8  \columnwidth]{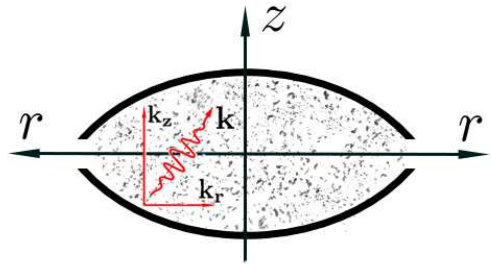}}
\caption{
 A scheme of the optical microcavity filled with the organic dye to observe Bose-Einstein condensation of photons. \cite{Klaers1,Klaers2}. Due to the symmetry of the sytem, the quantum state of a photon is described by the longitudinal $k_z$ and transverse $k_r$ wave numbers. 
}
\label{spectrum}
\end{figure}

We should also emphasize  that the system under consideration is effectively two-dimensional (2D). Talking about the light condensation in the system  \cite{Klaers1,Klaers2} one implies that photons condense in the 2D momentum space $\left\{  \mathbf{k}_{\mathbf{r}} \right\}$.  The condensation of photons thus means the occupation of the subspace  $\left\{  \mathbf{k}_{\mathbf{r}} = 0 \right\}$. The phenomenon takes place only after the total amount of free photons in the system exceeds the critical number

\begin{equation}\begin{split}\begin{gathered}
\label{T^2}
N_c \sim \left(   
\frac{T} {\hbar \Omega}
\right)^2,
\end{gathered}\end{split}\end{equation}

\no
which is mesoscopical for the current experimental conditions  \cite{Klaers2}, $N_c \sim 10^5$.
Note that the $T^2$ dependence here is the consequence of the effectively 2D geometry, and was justified in the experiments \cite{Klaers2}.

\section{Radiation spectrum of collisionless photon gas with condensate}

On this step we concentrate on the mechanism of radiation inside a cavity. Consider a photon with energy $\hbar \tilde \omega$, which at time $t=0$ was emitted by a molecule. To describe the radiation of this single photon during its lifetime (i.e. before it will be re-absorbed by another molecule), we can consider the single photon field between two collisions with medium structure units to be of a form $\xi=\xi_0  e^{i \tilde{\omega} t}$.
Therefore, in a general case, the 
amplitude of a signal emitted from the one photon during the time $t$ as it travels without collisions, is given by expression:

\begin{equation}\begin{split}\begin{gathered}
\label{amplitude}
A\left( \omega; \tilde{\omega} ,t \right)=\frac{{{\xi }_{0}}}{2\pi }\int\limits_{0}^{t}{{{e}^{i \tilde\omega t}}{{e}^{-i  \omega t}}dt}
\\
=\frac{{{\xi }_{0}}}{\pi }{{e}^{i\left(\tilde \omega - \omega  \right)t/2}}\frac{\sin \left( \omega -{\tilde \omega} \right)t/2}{\left( \omega -{\tilde \omega } \right)}.
\end{gathered}\end{split}\end{equation}

\no
From now we drop all constant prefactors, because we are interested only in the shape of the spectrum (recall that the current experimental data  for intensity \cite{Klaers2} is given  in arbitrary units only).
Apparently, the intensity of a signal emitted by a photon as it travels between two consequent collisions can be calculated in a conventional way:

\begin{equation}\begin{split}\begin{gathered}
\label{signal intensity}
I\left( \omega ; {\tilde \omega },t \right)
\propto
{{\left| A\left( \omega ; {\tilde \omega },t \right) \right|}^{2}}
\propto
\frac{{{\sin }^{2}}\left( \omega -{\tilde \omega } \right)t/2}{{{\left( \omega -{\tilde \omega } \right)}^{2}}}.
\end{gathered}\end{split}\end{equation}

\no
Now we need to take into account the finite lifetime of a photon in the cavity filled with absorbing medium. In other words, one should consider that the probability  for a photon to be free vanishes exponentially as time passes. 
Due to the casual and independent character of events when a photon with a given energy vanishes due to interactions with structural units of medium, the time between two consequent events follows the Poisson statistics. If we introduce the average lifetime 
$\tau$ of a photon in a cavity with given medium, the probability of an event, when the photon may be absorbed during the time interval $[t_{0},t_{0}+dt_{0}]$, can be written as:

\begin{equation}\begin{split}\begin{gathered}
\label{probability}
{{\left. dp \right|}_{\left[ {{t}_{0}},{{t}_{0}}+d{{t}_{0}} \right]}}
=\frac{1}{\tau }{{e}^{-{{t}_{0}}/\tau }}d{{t}_{0}}.
\end{gathered}\end{split}\end{equation}

\no
Of course, such a probability distribution is in some sense artificial: It does not correctly  take into account events, which are happening during time $t \sim l/c$, where $l$ is a typical distance between medium particles (molecules or atoms) and $c$ is the speed of light. In other words, the maximum of absorption probability is artificially shifted to the time $t=0$. However, from the physical considerations it is clear that absorption probability should start from zero: Immediately after a photon has been emitted by one molecule (or atom), there is no chance of absorption by other molecule (or atom). The validity of the approach used  for the expression  \eqref{probability} is justified by the strong time hierarchy, $ \tau \gg l/c \sim 0$. 
Of course, one can use more exact probability distributions, but if we want to calculate the shape of spectrum analytically, Eq.\eqref{probability} is a "necessary evil".

Now we actually proceed to the time averaging of signals.
According to Eq.\eqref{probability}, the average intensity radiated by a photon for the small interval of time $dt_{0}$ is expressed as:

\begin{equation}\begin{split}\begin{gathered}
{{\left. dI\left( \omega ; {\tilde \omega } \right) \right|}_{\left[ {{t}_{0}},{{t}_{0}}+d{{t}_{0}} \right]}}
=\frac{1}{\tau }I\left( \omega ; {\tilde \omega },{{t}_{0}} \right){{e}^{-{{t}_{0}}/\tau }}d{{t}_{0}}.
\end{gathered}\end{split}\end{equation}

\no
Now, integrating over all the  possible times a photon could live from emission to re-absorption,  $0<t_{0} < \infty $,  and taking into account Eq.\eqref{signal intensity}, one obtains the time-averaged signal intensity $I (\omega; \tilde \omega)$ that origins form a single photon with energy $ \hbar \tilde \omega$ [see \eqref{dispersion}]:

\begin{equation}\begin{split}\begin{gathered}
\label{spect lorentz}
I\left( {\omega ; \tilde \omega} \right)
 = \int
{{\left. dI\left( \omega ; {\tilde \omega } \right) \right|}_{\left[ {{t}_{0}},{{t}_{0}}+d{{t}_{0}} \right]}}
\\
 = \int\limits_0^\infty  {\frac{{d{t_0}}}{\tau }I\left( {\omega ; \tilde \omega , {t_0}} \right){e^{ - {t_0}/\tau }}}  
\propto
 \frac{1}{{{{\left( {\omega  -\tilde \omega} \right)}^2} + {\tau ^{ - 2}}}}.
\end{gathered}\end{split}\end{equation}

The crucial peculiarity of the system under study is that photons inside a cavity are in the  thermodynamical equilibrium with thermostatic medium. Therefore, the behavior of the equilibrium system obeys Bose-Einstein statistics with non-vanishing chemical potential of photons (see e.g. \cite{Klaers2,Kruchkov,Kruchkov1}). To calculate the radiation spectrum of the whole system $I ( \omega )$ , one therefore needs to average the single-photon signal intensity $I (\omega; \tilde \omega)$ over all possible states of photons in  the system, including a Bose-Einstein condensed phase. One should however be accurate to account correctly a contribution of the Bose-Einstein-condensed  light. To do it, we use the result, obtained by Kruchkov \cite{Kruchkov1}  for the  spectral density of free photons $\nu_{\mathbf{k}_{r}}$ in the system with Bose-Einstein-condensed light, which is normalized on the total number of free photons in the system, $\int{{{\nu }_{{{\mathbf{k}}_{r}}}}d{{\mathbf{k}}_{r}}}=N$. For  systems with geometry close to that one used in current experiments \cite{Klaers1,Klaers2}, this quantity can be written as \cite{Kruchkov1}:

\begin{equation}\begin{split}\begin{gathered}
\label{spectral density}
{{\nu }_{{{\mathbf{k}}_{r}}}}=
{{N}_{{{\mathbf{k}}_{r}}=0}} \
\delta \left( {{\mathbf{k}}_{r}} \right)
\\
+\frac{{{g}^{*}}q}{4 \pi}\frac{T}{{{m}^{*}}{{\Omega }^{2}}} \, \ln \left[ 1-\exp \left( -\frac{{{\hbar }^{2}}k_{r}^{2}}{2{{m}^{*}}T} \right) \right]^{-1},
\end{gathered}\end{split}\end{equation}

\no
where  $\delta \left( {{\mathbf{k}}_{r}} \right)$ is the two-dimensional Dirac delta-function, and ${{N}_{{{\mathbf{k}}_{r}}=0}}$ is a number of condensed photons in the system with temperature $T$. Dimensionless constants $g^{*}$  and $q$ are correspondingly  the effective  degeneracy of photon's energy in a given system (see  \cite{Kruchkov,Kruchkov1}) and longitudinal mode number (for instance, $q=7$ in experiments of Ref.\cite{Klaers2}).
Note that the number of condensed photons ${{N}_{{{\mathbf{k}}_{r}}=0}}$ beneath the phase transition point depends both on the temperature of the system $T$, its geometry, and also on properties of medium, which was used to thermalize the light \cite{Kruchkov1}:

\begin{equation}\begin{split}\begin{gathered}
\label{condensed}
{{N}_{{{\mathbf{k}}_{r}}=0}}(T<T_{c})
=
N_{\gamma}-
\frac{{{\pi }^{2}}}{12}{{g}^{*}}q{{\left( \frac{T}{\hbar \Omega } \right)}^{2}}
\\
- N_{a} \left( 1+\frac{{{g}_{{{\alpha }_{1}}}}}{{{g}_{{{\alpha }_{2}}}}}{{\operatorname{e}}^{\Delta /T}} \right)^{-1},
\end{gathered}\end{split}\end{equation}

\no 
where $N_{\gamma}$ is a total amount of photons, pumped into the system, and quantities $g_{\alpha_{1}}$,$g_{\alpha_{2}}$, $N_a$, $\Delta$ describe the properties of in-cavity medium in two-level model approximation (for details see \cite{Kruchkov1}; for validity of two-level model see also \cite{Klaers1,Klaers2,Sobyanin}). 
We emphasize here one more time that the system under study is effectively two-dimensional; in particular, this was taken into account in the expression \eqref{spectral density} for 2D spectral density of photons.

Now one should  take into account the following circumstance. Gaining the Bose-Einstein condensed state, photons are loosing their energy down to the lowest possible energy level in the system, $\hbar \tilde \omega = \hbar \omega_{0}$. Comparing this statement with expression \eqref{dispersion}, one notes that all the condensate is therefore  localized only in the center of the optical  cavity, $r \approx 0$. The spatial extent of the condensed cloud is  finite due to both quantum fluctuations and thermal fluctuations, $\left\langle r \right\rangle  \sim {10^{ - 5}}m$ (for experimental data see \cite{Klaers2}, for theoretical calculations see \cite{Kruchkov1}). However, this quantity is negligibly small comparing to a size of the cavity, and as a result Klaers \etal  measured spectral characteristics by placing a spectrometer  centrally with respect to the  cavity \cite{Klaers2}. In the vicinity of a cavity center, the dispersion relation of photons \eqref{dispersion} obtains a form:

\begin{equation}\begin{split}\begin{gathered}
\label{dispersion at r=0}
\hbar  \tilde{\omega} \left( k_{r},r \approx 0 \right) \approx  \hbar {\omega _0} + \frac{{{\hbar ^2}k_r^2}}{{2{m^*}}}.
\end{gathered}\end{split}\end{equation}

\no
Therefore, the statistically averaged radiation spectrum $I ( \omega) $, measured by a spectrometer, 
rises as a result of averaging over all possible 2D momenta of photons  in a cavity with medium.
To do it, one needs to integrate  Eq. \eqref{spect lorentz} taking into account expression \eqref{dispersion at r=0},

\begin{equation}\begin{split}\begin{gathered}
\label{radiation spectrum def}
I\left( \omega  \right) \propto \int\limits_{{{\bf{k}}_r}} {\frac{{{\nu _{{{\bf{k}}_r}}}d{{\bf{k}}_r}}}{{{{\left( {\hbar \omega  - \hbar {\omega _0} - {\hbar ^2}k_r^2/2{m^{*}}} \right)}^2} + {\gamma ^2}}}},
\end{gathered}\end{split}\end{equation}

\no
where we have introduced $\gamma= \hbar  \tau^{-1}$ that is  the average photon absorption rate.
Substituting the implicit expression for spectral density of photons in a cavity  \eqref{spectral density}, and introducing a new dimensionless variable $\xi  = {\hbar ^2}k_r^2/2{m^*}T$, one can easily obtain from \eqref{radiation spectrum def} the expression for radiation spectrum $I (\omega)$ of photons trapped inside a cavity:

\begin{equation}\begin{split}\begin{gathered}
\label{radiation spectrum}
I\left( \omega  \right) \propto \frac{{{N_{{{\bf{k}}_r} = 0}}}}{{{{\left( {\hbar \omega  - \hbar {\omega _0}} \right)}^2} + {\gamma ^2}}}
\\
+
\frac{{{g^*}q}}{{2 }}{\left( {\frac{T}{{\hbar \Omega }}} \right)^2}\int\limits_0^\infty  {\frac{{d\xi \,\ln {{\left[ {1 - {e^{ - \xi }}} \right]}^{ - 1}}}}{{{{\left( {\hbar \omega  - \hbar {\omega _0} - T\xi } \right)}^2} + {\gamma ^2}}}} .
\end{gathered}\end{split}\end{equation}

\no
The obtained expression contains an interesting prefactor before the integral. One can notice that this prefactor is proportional to the number of non-condensed photons in the system under study
 (see \cite{Kruchkov1}):

\begin{equation}\begin{split}\begin{gathered}
\label{non-condensed}
{{N}_{{{\mathbf{k}}_{r}} \ne 0}}
=
\frac{{{\pi }^{2}}}{12}{{g}^{*}}q{{\left( \frac{T}{\hbar \Omega } \right)}^{2}}.
\end{gathered}\end{split}\end{equation}

\no
Therefore the expression \eqref{radiation spectrum} possesses an interesting symmetry between the condensed phase $N_{{\mathbf{k}}_{r}=0}$ and non-condensed phase $N_{{\mathbf{k}}_{r} \ne 0}$. Introducing now Eq.\eqref{non-condensed}, one can significantly simplify the expression for the radiation spectrum $I (\omega )$:

\begin{equation}\begin{split}\begin{gathered}
\label{spectrum}
I\left( \omega  \right) \propto \frac{{{N_{{{\bf{k}}_r} = 0}}}}{{{{\left( {\hbar \omega  - \hbar {\omega _0}} \right)}^2} + {\gamma ^2}}}
 +
 \frac{{{N_{{{\bf{k}}_r} \ne 0}}}}{{{T^2}}}f\left( {\frac{{\hbar \left( {\omega  - {\omega _0}} \right)}}{T};
 {\frac{\gamma }{T}}
   } \right).
\end{gathered}\end{split}\end{equation}

\no
where $f (\alpha ; \beta )$ is a function of dimensionless parameters $\alpha$,$\beta$:

\begin{equation}\begin{split}\begin{gathered}
\label{thermal function}
f\left( {\alpha ;\beta } \right) = \frac{{{\pi ^2}}}{6}\int\limits_0^\infty  {\frac{{d\xi \,\ln {{\left[ {1 - {e^{ - \xi }}} \right]}^{ - 1}}}}{{{{\left( {\xi  - \alpha } \right)}^2} + {\beta ^2}}}}.
\end{gathered}\end{split}\end{equation}

\no
Expressions \eqref{spectrum} and \eqref{thermal function} fully describe the radiation spectrum $I (\omega )$ (intensity as a function of measured frequency) of Bose-Einstein-condensed light in the approximation of constant lifetime of photons $\tau$.
Note that the first term in expression \eqref{spectrum} is the contribution of the condensed photons, i.e. it describes the influence of the ground state occupation on the result spectrum,  and the second term is the contribution of non-condensed photons. Both of these contributions are important.
Note also, that in the case when almost all photons are in condensate, the shape of spectrum near $\omega = \omega_{0}$ is close to lorentzian curve.

One should make here a following remark. In principle, the lifetime of a photon in a cavity with medium can depend on it's frequency, $\tau = \tau ( \tilde \omega) $. Therefore, formula \eqref{spectrum} that defines the radiation spectrum of photons in a cavity can be modified. The frequency dependence of photon lifetimes is nevertheless a non-trivial and complicated  issue  that is beyond the topic of the present letter. 
However, one could consider that in the simplest model the photon lifetime is significantly different in the narrow vicinity of the cutoff frequency $\omega_0$,

\begin{equation}\begin{split}\begin{gathered}
\label{lifetime}
\tau (  \tilde \omega) =
\begin{cases}
   \tau_{0},& \text{if} \ \    \tilde \omega \approx \omega_0,
\\
    \tau, & \text{else,}
\end{cases}
\end{gathered}\end{split}\end{equation}

\no
Consequently, one needs to take this circumstance into account performing integration in expression \eqref{radiation spectrum def}.
Due to the structure of the integrand, 
the frequency $\omega_0$ is  treated in the integral \eqref{radiation spectrum def} apart from other frequencies; in the main approximation one therefore obtains:

\begin{equation}\begin{split}\begin{gathered}
\label{spectrum1}
I\left( \omega  \right) \propto \frac{{{N_{{{\bf{k}}_r} = 0}}}}{{{{\left( {\hbar \omega  - \hbar {\omega _0}} \right)}^2} + {(\hbar / \tau_{0} )^2}}}
 \\
 +
 \frac{{{N_{{{\bf{k}}_r} \ne 0}}}}{{{T^2}}}f\left( {{\hbar \left( {\omega  - {\omega _0}} \right)}/T;
 \hbar \tau^{-1} / T
   } \right).
\end{gathered}\end{split}\end{equation}

\no
Formula \eqref{spectrum1} is the final formula we will use in couple with \eqref{thermal function} to compare with experimental results.
Also, one should always keep in mind that there can exist a weak background from non-thermalized photons.

\begin{figure}[t]
\center{\includegraphics[width=1.0 \columnwidth]{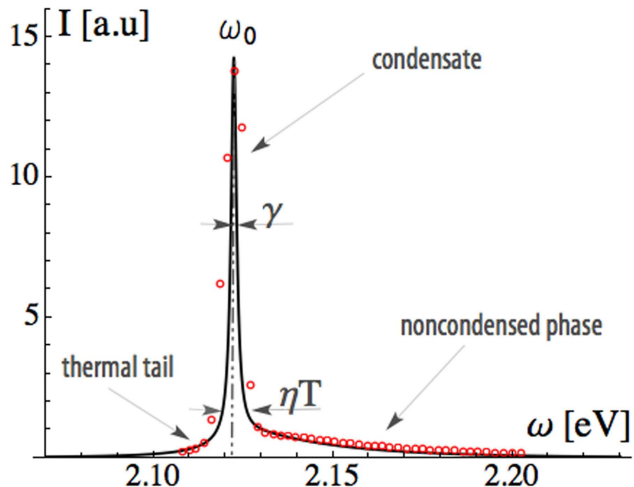}}
\caption{
Radiation spectrum of a cavity with condensed light: experiment and theory. Experimental points were extracted from Ref.\cite{Klaers2}, theoretical curve plotted taking into account \eqref{spectrum1} and \eqref{thermal function}. The condensate peak corresponds to the cutoff frequency $\omega_0$ and can be characterized by the width $\gamma$ and the thermal blur $\eta T$, where $\eta$ is a fraction of noncondensed photons. A thermal tail in a "forbidden" energy range [see \eqref{dispersion}] exists because the system is described by a finite temperature $T$. A noncondensed phase is always present for $T \ne 0$.
}
\label{spectrum}
\end{figure}

We  show the comparison between the presented phenomenological model and experiment (see Fig.2). Experimental points were extracted from the Ref.\cite{Klaers2}, where the intensity $I (\lambda )$ was given in arbitrary units (a.u.) as a function of wavelength $\lambda = 2 \pi c/ \omega$; theoretical curve was plotted using the derived equations for $I(\omega)$, \eqref{spectrum1} and \eqref{thermal function}. Parameters of the system were  taken from \cite{Klaers2}. The condensate peak can be described by the photon absorption rate $\gamma ( \tau_0 ) = \hbar/ \tau_0$  and the thermal blur that is proportional to the temperature of the system $T$ and an amount of noncondensed photons (see also Ref.\cite{Kruchkov1}). In the energy forbidden area, i.e. where $\tilde \omega < \omega_0$ [see \eqref{dispersion}], one can observe a thermal tail, caused by the finiteness of photon lifetimes and the significant role of thermal fluctuations on the description parameters of the system (see also \cite{Kruchkov1}).
 The spectral width $\gamma ( \tau_0 )$ was estimated to be no more than the spectrometer resolution (see also Conclusion and Ref.\cite{Klaers2}).
The present phenomenological model fits good the current experiments.

\section{Influence of photon-photon interactions}

Photons in the system under study were shown to interact weakly \cite{Klaers2}. Therefore after considering the non-interacting theory, it is naturally to take these interactions into consideration.
 However, as we have already mentioned, the light condensate differs from the case of atomic BECs,  and consequently the techniques used in the theory of ultracold atomic condensates may not be applicable and should be used very carefully. 
The main delicacy here is the "hot" temperature of the light condensate, $T=300$ K.
 In the general case for a finite-temperature condensates one should better use Peletminskii-Kirkpatrick equations \cite{Peletminskii,Kirkpatrick} or similar.
The authors of the original experiment \cite{Klaers2}, however, use the so called Gross-Pitaevskii  equation \cite{Gross,Pitaevskii}. Physical approximations, laying in the basement of the Gross-Pitaevskii equation, includes the two-body contact interactions and neglecting the anomalous contributions to self-energy \cite{Beliaev}.
These assumptions are valid mostly for the dilute 3D Bose-Einstein condensates near $T \approx 0$.
In contrast, in our case \cite{Klaers2} one has to deal with the 2D Bose-Einstein condensate at a room temperature.
The use of Gross-Pitaevskii equation may be justified, for example, for the particular case when most of photons are condensed.
Following the authors of the original paper \cite{Klaers2}, for the sake of simplicity of numerical estimates,  in this section we use the Gross-Pitaevskii model,

\begin{equation}\begin{split}\begin{gathered}
\label{GP}
\left\{ 
- \frac{ \hbar^2}{2 m^*} \Delta +\frac{1}{2}m^* \Omega^2 r^2 - g \left| \psi \left( r \right)  \right| ^ 2
\right\}
\psi \left( r \right)
=
\mu \, \psi \left( r \right), 
\end{gathered}\end{split}\end{equation}

\no
where the nonlinear term $g \, \psi^2 \left( r \right)  \psi^* \left( r \right) $ reflects the strength of interactions in $\delta$-like pseudopotential model of two-body collisions.
The dimensionless interaction parameter $\tilde g \equiv \left(m^* /\hbar^2 \right) g $ in the system under study was reported to be $\tilde g = 7 \times 10 ^{-4}$ (see \cite{Klaers2}). Consequently, in the interesting us region $r \approx 0$ the photon-photon interactions redefine a photon's energy by the amount

\begin{equation}\begin{split}\begin{gathered}
\label{interaction energy}
\delta \varepsilon ^{(int)}  \approx \tilde g \frac{\hbar^2 \, n (  r \approx 0  )}{m^*}.
\end{gathered}\end{split}\end{equation}

\no
The density of photons in the center of condensate $n \left( r \approx 0 \right) \equiv n_0$ can be extracted directly from the experimental measurements \cite{Klaers2}, $n_0 \sim 5 \times 10^{12} \, m^{-2}$. 
Substituting now the effective mass of a photon $m^* \approx 7 \times 10^{-36}$ kg 
\cite{Klaers2}
and the above mentioned physical quantities, one finds the interaction energy correction for a photon in the system under study,

\begin{equation}\begin{split}\begin{gathered}
\label{interaction energy}
\delta \varepsilon ^{(int)}  \sim 10^{-5} \, eV.
\end{gathered}\end{split}\end{equation}

\no
Note that this quantity is very small: it is two orders of magnitude smaller than the spectrometer resolution ($\sim 10^{-3} \, eV$) and three orders smaller than the temperature fluctuations of the noncondensed photons in the system 
($\sim 10^{-2} \, eV$). We recall that the typical energy of a photon in the system is $\hbar \omega_0 \approx 2 \, eV$.

Now imagine a daemon, who knows the exact dependence $\tau \left( \omega \right)$ for the system in the non-interacting case. The daemon observes a photon with energy $2 \, eV$ coming from the region with no light condensate, and registers the change of its energy of order $10^{-5} \, eV$ caused by interactions with other photons. This second-quantization daemon considers that the photon with energy $2 \, eV$ died and the new photon with slightly smaller energy was born, so he  carefully re-calculates the energy-dependence of photons lifetimes, $\tau \left( \omega \right) \to \tilde \tau \left( \omega \right)$. However, his friend Lucifer sitting inside the spectrometer with resolution $10^{-3} \, eV$ neglects the renormalization in $\tilde \tau \left( \omega \right)$, which is of order $\delta \varepsilon ^{(int)}/T \sim 0.1 \% $, and reveals for the observer no more than the  information blurred by thermal fluctuations, without any hint on photon-photon interactions.

\section{Conclusion}

In the present paper we have proposed a rather simple phenomenological model, explaining the shape of spectrum emitted by a photon gas with condensate. 
The main assumption of the model was the quasi-energy-independent  lifetime of photons and the assumption of casual and independent character of vanishing of a photon with the given energy, which leads to the lorentzian broadening of a particular line profile. The lorentzian broadening of an energy level is not a new result and should be considered here only as a technique to link in a non-contradictory way the emitted electromagnetic intensity at the frequency $\omega$ with the momentum averaging \eqref{radiation spectrum def}, which takes into account both the thermal equilibrium of the photons and the presence of condensate. This is a rather simple but elegant method, which were not used in the previous study \cite{Klaers2}. In particular, in the Ref.\cite{Klaers2}, the intensity of of light was calculated on the basis of Bose-Einstein distribution, without the direct consideration of the condensed fraction. The proposed in the present paper phenomenological model allows to obtain the new   theoretical results:

\no
(i). The thermal tail in the ``forbidden'' energy part of spectrum rises due to the influence on noncondensed photon and depends in the temperature of the system and other parameters, and therefore can be suppressed.

\no
(ii) The height of the condensate peak is {\it{linearly proportional}} to the number of photons in condensate.

\no
(iii) Meanwhile, the width of the condensate peak is blurred by the ``temperature'' fluctuations of the non-condensed photons, and by choosing the appropriate parameters (temperature, condensate fraction etc. ) this blur can be also  suppressed (see also \cite{Kruchkov1}).

The role of photon-photon interactions, considered in the Sec.IV, was shown to be neglectable for the current set of experimental parameters. This happens essentially because of the smallness of photon-photon interaction parameter 
and low 2D density of condensed photons in the system. 

One of the crucial aspects that should be clarified in the further research is the energy-dependence of the photons lifetimes $\tau ( \omega )$. 
As we have already mentioned, 
in the present paper 
we consider lifetimes of photons to be quasi-independent on their energies \eqref{lifetime}, which can be reasonable in the present system \cite{Klaers1,Klaers2} where due to  the specifics in the thermalization process the lifetime is sufficiently different only for photons that are close to the energy of 2 eV \cite{Klaers1}.
Despite the seeming modesty of the used approach, it should be mentioned that due to the specifics of the integral \eqref{radiation spectrum def} the main result will be similar to the form of \eqref{spectrum1} even in the case with exactly known dependence $\tau ( \omega )$.
Note also, that the original paper \cite{Klaers2} operates with the constant (energy-independent) lifetime of photons in the microcavity that in turn leads to the reliable results \cite{Klaers2}. However, the author of the present paper agrees that there are multiple factors which cause photons with alternate energies to exist the different amount of time. Among these factors are different absorption/emission probabilities of photons with alternate energies by the organic dye, the influence of cavity modes, small  imperfections of mirrors and multiple scattering processes which take place during the thermalization stage of kinetic evolution (see Refs.\cite{Klaers1,Klaers2,Kruchkov1,Sobyanin,Kirton}). And if we can deal with the first listed issues, the exact physics of inelastic scattering  of light on the organic dye molecules currently is not fully understood. 
This is a separate and nontrivial problem that is far beyond the scope of the present paper. The author hopes to address the issue in future.

Finally, one should also make  the following remark. The physical quantity $\gamma$, shown on the Fig. 2, in the case of the real experiment \cite{Klaers2} does not depend on photon lifetimes, but is determined by the resolution of the spectrometer $\Delta \varepsilon^{(res)} $.
The true lineshape near the condensation frequency $\omega = \omega_0$ can be even narrower, but we cannot measure this experimentally. In such a way we have only estimated the lowest boundary for the lifetime $\tau_0$, $\tau_0 \ge  \hbar / \Delta \varepsilon^{(res)} $.
In this sense, the author asks to consider the narrow part of the spectrum near $\omega \approx \omega_0$ only as a desirable fit.

The study do not claim to be the exact and comprehensive theory, and it could and should be improved.

\

Acknowledgments.  The work was supported by Swiss National Science Foundation (SNSF) and its Sinergia network MPBH. 
I would like to thank Igor Vakulchik for his valuable comments during the preparation of the paper.

\end{document}